\documentclass{article}
\usepackage{graphics}

\newcommand{\ket}[1]{{|#1\rangle}}

\author{Christof Zalka \\ zalka@t6-serv.lanl.gov}

\title{A Grover-based quantum search of optimal order for an unknown
number of marked elements}

\begin{document}
\maketitle
\begin{abstract}
We want to find a marked element out of a black box containing $N$
elements. When the number of marked elements is known this can be done
elegantly with Grover's algorithm, a variant of which even gives a
correct result with certainty. On the other hand, when the number of
marked elements is not known the problem becomes more difficult. For
every prescribed success probability I give an algorithm consisting of
several runs of Grover's algorithm that matches a recent bound
\cite{buhrman} on the order of the number of queries to the black
box. The improvement in the order over a previously known algorithm is
small and the number of queries can clearly still be reduced by a
constant factor.
\end{abstract}

\section{Introduction}
We are given a subroutine which for every one of $N$ possible inputs
gives us a $0$ or a $1$ as output. The subroutine is in a ``black
box'', so we are not allowed to see what algorithm it uses. The
problem is to find an input for which the subroutine gives $1$ (a
``marked'' input or ``element'') or to say that there is no such
input.

Classically there is no other way than to try out many inputs on the
subroutine. On the other hand, if the subroutine also works for
superpositions of inputs

\begin{displaymath}
\sum_x c_x \ket{x} \quad \to \quad \sum_x c_x \ket{x,f(x)}
\end{displaymath}
we can use Grover's quantum searching algorithm \cite{grover} to
find a ``marked element'' (an input that yields $1$) which gives us a
quadratic speedup relative to the classical method. In particular
when we know the number $t$ of marked elements, Grover's algorithm
will find one of these elements with high probability using only about
$\pi/4 \sqrt{N/t}$ queries to the subroutine. A slight modification of
this algorithm can actually increase the success probability to $1$
while using at most one more query (see below). All this also works
when we are only told that the number of marked elements is either $t$
or $0$.

\section{Unknown number of marked elements}
On the other hand, when we don't know the number of marked elements,
Grover's algorithm doesn't perform that well. In particular to get a
certainly correct answer any algorithm will use on the order of $N$
queries as for the classical case \cite{buhrman}.

If we allow for some error probability, a Grover type quantum search
still helps. Note that when the quantum algorithm outputs a candidate
marked element, we can check it with just one more query. Thus the
only error that can happen, is that the algorithm wrongly says that
there are no marked elements. Thus the error is so-called one sided
and we have not only a quantum Monte Carlo algorithm (BQP) but
actually a quantum-RP algorithm.

So the problem now is: Given $N$ and some upper bound
on the error probability $\epsilon$, find an algorithm that uses as
few queries to the black box as possible. Note that by error
probability we mean the error probability for the worst case, which
here in particular means the number of marked elements for which the
algorithm performs worst.

\subsection{A simple algorithm}
A simple, but not optimal, solution to this problem is to run Grover's
algorithm many times for some random number of iterations (= number of
queries) between $0$ and $\pi/4 \sqrt{N}$, which is about the optimum
for just one marked element. We choose the number of iterations of
each run uniformly at random from the given range. From the evolution
of the state vector (of the QC) in Grover's algorithm (see below) it
is easy to see that in each such run and for any number of marked
elements (except $0$) the probability of finding a marked element is
about $1/2$. Actually a careful analysis (see below) shows that for
the worst-case number of marked elements $t$, it's about $0.3914$. By
repeating this many times we get asymptotically at most about the
following number of queries:

\begin{displaymath}
T \approx \pi/4 \sqrt{N} \frac{\ln(\epsilon)}{\ln(1-0.3914)}
\end{displaymath}

\subsection{The improved algorithm}
The proposed algorithm again basically consists of many runs of
Grover's algorithm, each for some numbers of iterations (=number of
queries). It consists of 2 parts plus possibly a third one to improve
it a bit. 

The first part checks for all numbers of marked elements from $1$ to
some maximum $t_0$, each time running Grover's algorithm for the
appropriate number of iterations, which is about $\pi/4 \sqrt{N/t}$
for $t$ marked elements. If the maximal error probability $\epsilon$
which we allow is below about $1/N$ we have to use the exact version
of Grover which is guaranteed to work for a given known number of
marked elements. The total number of queries of this first step is
about:

\begin{displaymath}
T \approx \sum_{t=1}^{t_0} \pi/4 \sqrt{N/t} \approx \pi/4 \cdot 2 \sqrt{N t_0}
\end{displaymath}
which I've obtained by approximating the sum with an integral. The
second part consists of a number of Grover runs for a random number of
iterations smaller than the last Grover run of the first part. Thus we
choose the number of iterations uniformly at random from the range
$0..\pi/4 \sqrt{N/t_0}$. This I propose to do $2 t_0$ times. Thus the
second part consists of at most the following number of queries:

\begin{displaymath}
T \approx 2 t_0~ \pi/4 \sqrt{N/t_0} = \pi/2 \sqrt{N t_0}
\end{displaymath}
It turns out that this choice of the number of Grover runs in the
second part which gives equal number of queries for the first and
second parts is optimal.

In the second part we have probability about $1/2$ to find a marked
element in every run, provided the number of marked elements is larger
than $t_0$. Say the actual (worst case) probability is some $p$ (which
is a bit smaller than 1/2 and which we will determine below), then
$\epsilon = (1-p)^{2 t_0}$ and we get the following number of queries
for the total algorithm as a function of $N$ and $\epsilon$:

\begin{equation} \label{res}
T \approx \pi \sqrt{N} \sqrt{\frac{\ln \epsilon}{2 \ln(1-p)}}
\end{equation}
Note the (admittedly small) improvement relative to the simpler
algorithm described above. Note also that $\epsilon$ as a function of
$T$ for a fixed $N$ goes as

\begin{displaymath}
\epsilon = O(e^{-c T^2})
\end{displaymath}
which is of course better than the exponential we could achieve by
just running the same probabilistic algorithm over and over.

It turns out that the worst case detection probability $p$ actually
occurs for a number of marked elements $t$ close to $N$. So if we want
we can add a third part to the algorithm where we take care of this
worst case by just classically checking a number of random inputs to
the black box. To achieve the maximum allowed error probability
$\epsilon$ this only takes a number of queries that is of a smaller
order than the number of queries used in the first two parts, thus we
can neglect the cost of this third part.

In the remainder of the paper I will determine the maximal error
probability $p$ for a Grover run with a random number of iterations
and I will also give a simple argument that an exact version of
Grover's algorithm can be constructed for a known number $t$ of marked
elements, which has been known before (I think it's by Peter Hoyer,
but can't find a reference).

\section{Finding the maximal error probability $p$}
First we have to review Grover's algorithm:

\subsection{Grover's algorithm for $t$ marked elements}
Each iteration of Grover's algorithm consists of the following four steps:

\begin{eqnarray*}
1. && \qquad \ket{y} \to -\ket{y} \qquad \mbox{for all marked elements} ~y \\
2. && \qquad H^l \\
3. && \qquad \ket{x} \to -\ket{x} \qquad \mbox{for all} \qquad x \not = 0 \\
4. && \qquad H^l
\end{eqnarray*}
The initial state is the uniform amplitude superposition of all
possible $N$ inputs to the black box. The first step involves querying
the oracle, while the second and fourth steps involve Hadamard
transforming each of the $l$ qubits (thus here we have $N=2^l$).

We can write the initial state as

\begin{displaymath}
\ket{\Psi_0} = \cos(\Theta_0) \frac{1}{\sqrt{N-t}} \sum_{x \not \in M} \ket{x}
+ \sin(\Theta_0) \frac{1}{\sqrt{t}} \sum_{y \in M} \ket{y}
\end{displaymath}
Where $M$ is the set of marked elements while $\cos(\Theta_0) =
\sqrt{N-t}/\sqrt{N}$ and $\sin(\Theta_0) = \sqrt{t}/\sqrt{N}$.

It turns out (and is easy to see) that after any number of
applications of the 4 steps of Grover's algorithm the state remains of
the above form, thus a superposition with real coefficients of the
uniform amplitude superposition of all unmarked (basis-) states and of
the uniform amplitude superposition of all marked states. The actual
calculation consists simply of applying the 4 above steps to each of
these 2 states. In every iteration the state vector gets rotated by
some angle $\Theta$, so after $n$ iterations we have:

\begin{displaymath}
\ket{\Psi_n} = \cos(\Theta_0+n \Theta) \frac{1}{\sqrt{N-t}} 
\sum_{x \not \in M} \ket{x}
+ \sin(\Theta_0+n \Theta) \frac{1}{\sqrt{t}} \sum_{y \in M} \ket{y}
\end{displaymath}
where $\Theta$ is given by

\begin{displaymath}
\cos(\Theta) = 1-2 \frac{t}{N} \qquad \mbox{and} \quad
\sin(\Theta) = 2 \frac{\sqrt{t}\sqrt{N-t}}{N}
\end{displaymath}
It is easy to check that $\Theta_0 = \Theta/2$ which will facilitate
the subsequent  calculations.

The probability of finding a marked element after $n$ steps is
$\sin^2 (\Theta_0 +n \Theta)$. If we choose the number of iterations
uniformly at random from the range $0..k-1$ we get the success
probability

\begin{displaymath}
p = \frac{1}{k} \sum_{n=0}^{k-1} \sin^2 (\Theta_0+ n \Theta) =
\frac{1}{2} \left( 1-\frac{\sin(2 k \Theta)}{2 k \sin(\Theta)} \right)
\end{displaymath}
Where the summation is easily accomplished because we can write the
trigonometric functions in terms of exponentials $e^{i...}$ which
gives us geometric series. Also we used $\Theta_0 = \Theta/2$. Note
that, as stated above, $k \approx \pi/4 \sqrt{N/t_0}$.

Now we have to look for the minimum of $p$ (=worst case) over the
range $t=t_0..N$ of marked elements. The lower end of this range
corresponds to $2 k \Theta =\pi$. The following plot shows $p$ as a
function of $\Theta$ (here for $k=7$). Note that $\Theta =0..\pi$
while $t=0..N$.

\resizebox{13cm}{8cm}{\includegraphics[18,144][588,566]{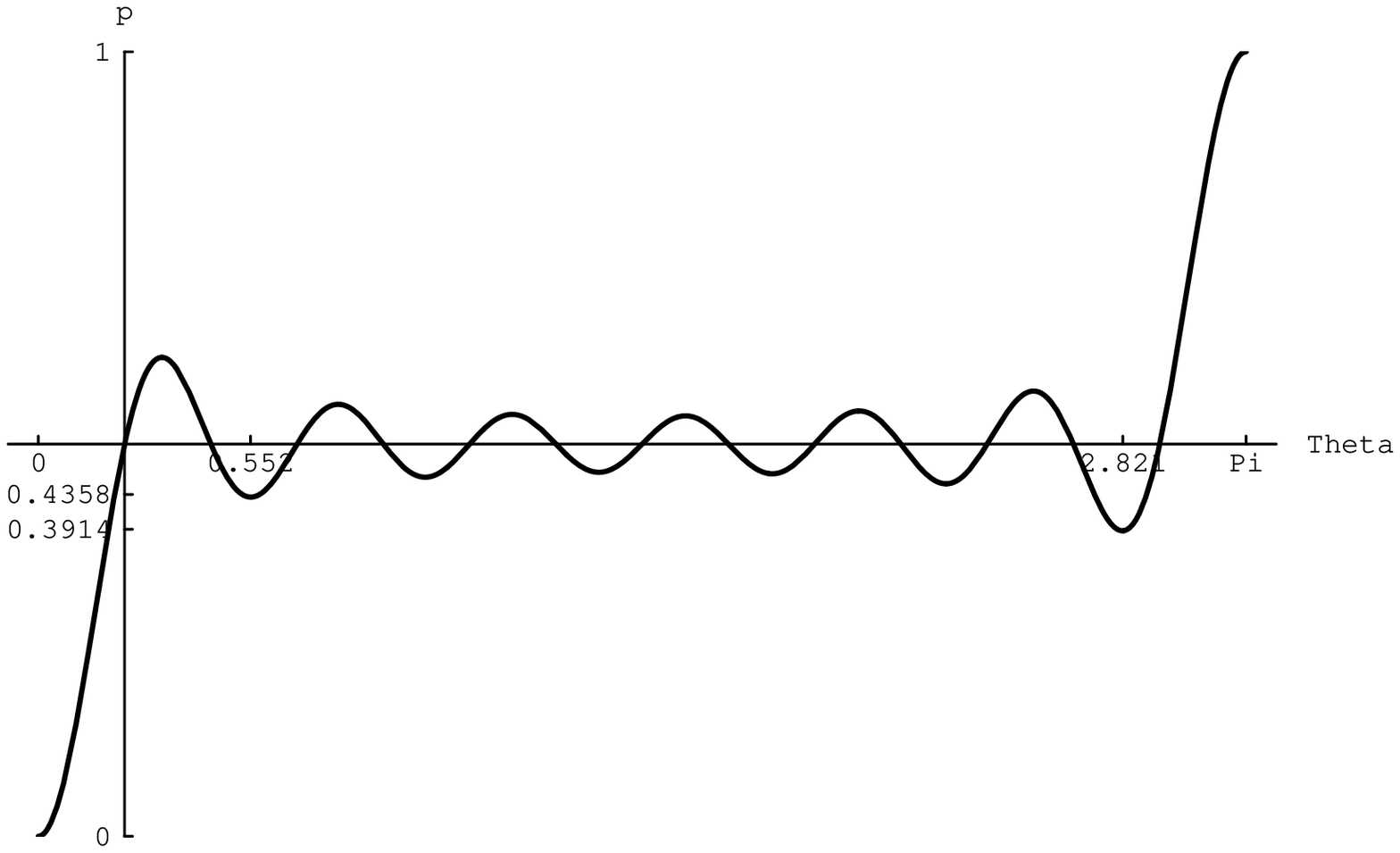}}

\resizebox{13cm}{8cm}{\includegraphics[18,144][588,566]{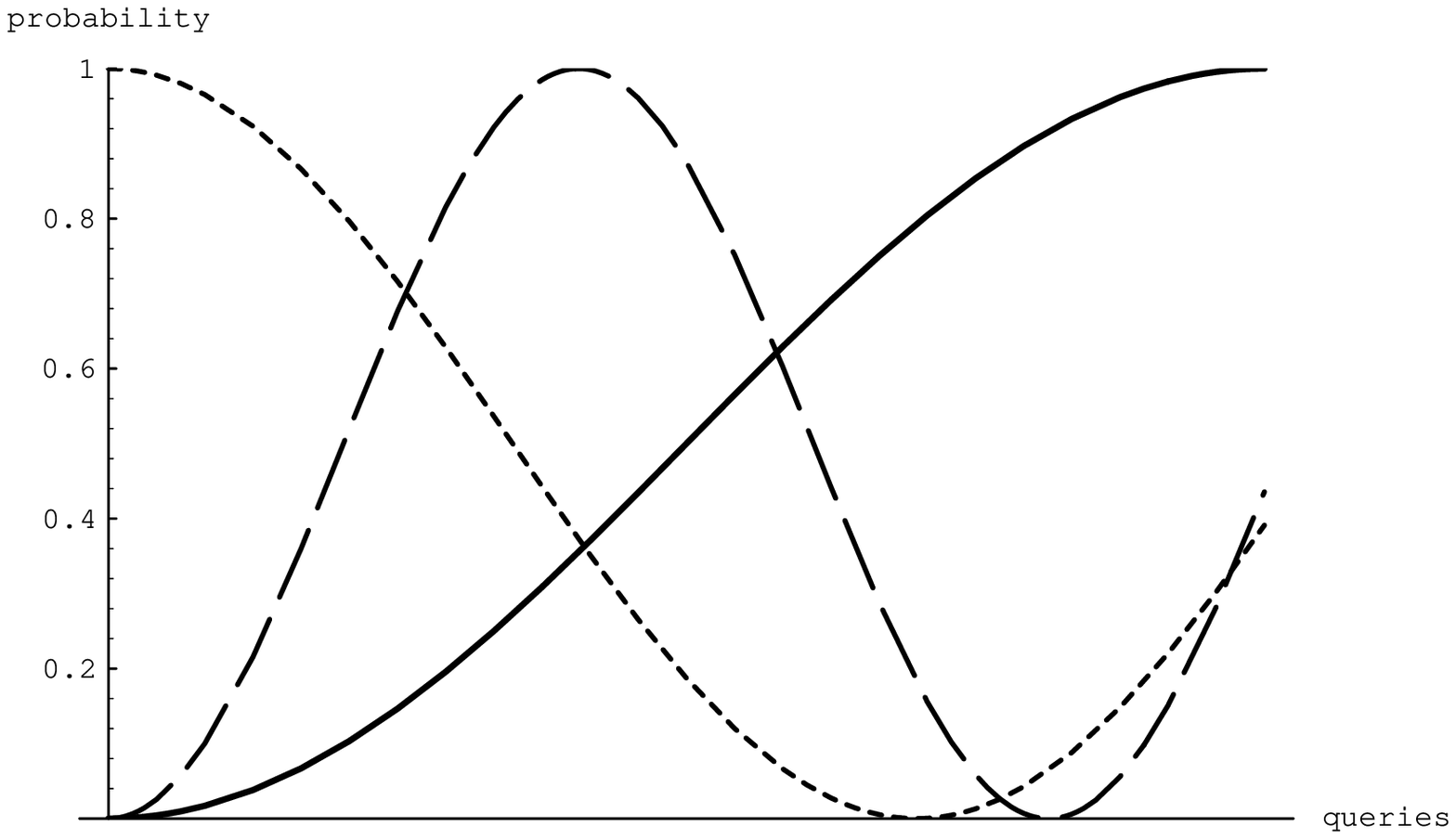}}

The point $t=t_0$ is where $p(\Theta)$ first reaches
$1/2$. Significant for us is the next local minimum $p_1 \approx
0.4358$ and the last local minimum $p_2 \approx 0.3914$, where I have
taken the limit $k \to \infty$ to get these values. These two values
are the absolute minimum (in our range) and the next larger local
minimum. If we include the third part of the algorithm (classical
checking) we don't have to worry about $p_2$, thus $p=p_1$. Then
asymptotically the number of queries becomes (from equation
\ref{res}):

\begin{displaymath}
T \approx 2.94 \sqrt{N} \sqrt{\ln 1/\epsilon}
\end{displaymath}

The second graph above shows how the success probability evolves in
Grover's algorithm for 3 cases of interest. The range shown in the
plot is the range which we use in the second part of the algorithm,
thus from 0 to about $\pi/4 \sqrt{N/t_0}$ iterations. The thick solid
line is for $t_0$ marked elements, for which the average probability
over this range is $1/2$. The dashed graph corresponds to the first
local minimum in the first plot, whereas the dotted graph gives the
last local minimum which occurs for only few unmarked elements.

\section{Exact Grover for known number of marked elements}
Here I give a simple argument why Grover's algorithm can easily be
modified to give a correct answer for a known number $t$ of marked
elements. The problem with standard Grover is that $\sin^2 (\Theta_0+n
\Theta)$ never exactly becomes $1$. Our task now is to modify the 4
steps in Grover's algorithm so that we get a smaller rotation angle
$\Theta' < \Theta$. Imagine we apply these (modified) 4 steps to the
initial (uniform amplitude) state. Usually in the 1. step we change
the phase of the marked states by $\pi$ which allows us to increase
the amplitude of the marked states to $\sin(\Theta_0+\Theta)$. If we
don't change the phase it is easy to see that the 4 steps of Grover's
algorithm don't change the initial state at all, thus the amplitude of
the marked states remains $\sin(\Theta_0)$. By continuity it is now
clear that we can adjust the absolute value of the amplitude of the
marked states to any value between these extremes. To get the
amplitude back to real and positive we then call the black box once
more to rotate the phase of the marked elements by the right amount.

Actually one can avoid this last (additional) call to the black box by
also choosing a different phase change in step 3, but this is not so
easy to explain. Of course there are also various other ways to make
Grover's algorithm exact.

\section{Remarks}
The algorithm we have constructed is clearly not optimal. It's order
{\it is} optimal, as stated in corollary 3 on page 6 of
\cite{buhrman}, but the performance can obviously still be improved by
a multiplicative constant. My guess is that the number of queries can
be reduced by at least a factor of $1.4$, but probably by a factor of
$2$ or more. Unfortunately no bound on the multiplicative constant is
specified in \cite{buhrman}.

In the second part of the algorithm we choose the number of iterations
of Grover at random. This is not very elegant and certainly not
optimal, but it makes the algorithm and the assessment of its
performance easy.

A general observation is that zero error algorithms are a rather
academic and unphysical concept. Any computer has some failure
probability, this is especially true for quantum computers.
Fortunately fault tolerant techniques allow us to greatly increase the
reliability. Essentially the error probability can be reduced
exponentially in the resources we invest into fault tolerance which
allows us to attain a reliability that is good enough for all
practical purposes. So what about my use of the exact variant of
Grover's algorithm? Actually we don't really need the exact version,
we merely need to be able to greatly reduce the error without using
many more queries. In a fault tolerant implementation we actually
anyways can't really apply phase rotations by any amount because we
only can use a finite ``universal'' set of gates. The better we want
to approximate a given phase rotation the more such gates we have to
use. Fortunately the number of gates necessary typically only
increases as the logarithm of the precision of the approximation.

I would like to thank Richard Cleve for telling me about the problem
which is solved in this paper.

\end{document}